\documentclass[aps,pra,floatfix,twocolumn,superscriptaddress]{revtex4-2}

\usepackage{amsmath}
\usepackage{amssymb}
\usepackage{subfigure}
\usepackage{graphicx}

\newcommand{\br}{\mathbf{r}}
\newcommand{\be}{\begin{eqnarray}}
\newcommand{\ee}{\end{eqnarray}}
\newcommand{\p}{\partial}

\def\ep#1{\langle #1 \rangle}

\begin{document}

\title{Larkin-Ovchinnikov-Fulde-Ferrell state of spin polarized atomic Fermi superfluid on a spherical surface}

\author{Yan He}
\affiliation{College of Physics, Sichuan University, Chengdu, Sichuan 610064, China}
\email{heyan$_$ctp@scu.edu.cn}

\author{Chih-Chun Chien}
\affiliation{Department of Physics, University of California, Merced, CA 95343, USA.}
\email{cchien5@ucmerced.edu}

\begin{abstract}
By implementing the Bogoliubov-de Gennes (BdG) formalism of population-imbalanced atomic Fermi gases with pairing interactions in a thin spherical shell, we characterize the Larkin-Ovchinnikov-Fulde-Ferrell (LOFF) state in such a compact geometry. We first construct a phase diagram showing where uniform solutions of spin-polarized Fermi superfluid from the BdG equation cease to exist due to the vanishing order parameter. Near the boundary, various LOFF states with spatially modulating order parameters and density profiles can survive as convergent solutions to the BdG equation. When both uniform and LOFF solutions are present, we compare their grand potentials to determine the energetically favorable state and find that the LOFF states with multiple nodes in the order parameter become more stable at higher spin polarization. However, the LOFF state only survives close to the phase boundary where the uniform solutions vanish, indicating fragility of the LOFF state on a spherical surface. We also briefly discuss possible implications.
\end{abstract}

\maketitle

\section{Introduction}
The Larkin-Ovchinnikov-Fulde-Ferrell (LOFF) state~\cite{LO64,FF64} reconciles pairing of fermions in Fermi superfluid and spin polarization by introducing modulating profiles of the order parameter and densities. In solid-state systems, possible evidence of the LOFF state has been reported~\cite{STEGLICH1996498,Imajo2022,Kinjo22,Wan2023}. Moreover, superconductor-ferromagnet heterostructures with proximity effects~\cite{Buzdin2005,Bergeret2005} may realize the LOFF state or other exotic states near the interface. More recently, ultracold atoms have emerged as a flexible platform for simulating many-body physics~\cite{Pethick-BEC,Ueda2010,Zwerger2012}. Meanwhile, intense search in low-dimensional ultracold atoms has also revealed possible hints of the elusive LOFF state~\cite{Liao2010,Revelle2016}. Meanwhile, plenty theoretical works on the LOFF state~\cite{PhysRevB.51.9074,doi:10.1143/JPSJ.76.051005,PhysRevB.73.214527,PhysRevB.75.014521,Kinnunen2018RPP} have elucidated many interesting aspects of this type of spin-polarized Fermi superfluid.

On the other hand, the realizations of spherical shell structures of ultracold atoms by the bubble trap in outer space~\cite{Carollo22,Lundblad_2023} or multi-species phase separation in a harmonic trap on earth~\cite{PhysRevLett.100.185301,PhysRevLett.97.060403,PhysRevLett.129.243402} have demonstrated interesting many-body physics in curved geometry exemplified by thin spherical shells. While theoretical analyses of bosonic atoms and Bose-Einstein condensation (BEC) in spherical shells are abundant~\cite{SphericalBECPRL19,SphericalSFPRL20,SphericalBECNJP20,PhysRevA.102.043305,PhysRevA.103.053306,PhysRevA.109.013301,PhysRevA.104.033318,PhysRevA.106.013309,PhysRevA.75.013611,PhysRevA.104.063310,PhysRevResearch.4.013122,He2023Soliton,Boegel_2023,PhysRevA.107.023319,PhysRevLett.132.026001,Arazo_2021,TononiReview23,TONONI20241}, there are relatively less studies of fermionic systems on a spherical surface. Ref.~\cite{FreeFermionShpere} studied free fermions and show discontinuous spectra due to the filling of orbital angular momentum states. Ref.~\cite{He22Sphere} lays the foundation of the mean-field theory of Fermi superfluid on a spherical surface across the BCS-BEC crossover. Spherical Fermi superfluids with topological excitations~\cite{He23Vortex} and spin-orbit coupling~\cite{He25SOC} have been characterized, and Ref.~\cite{Yan_RepulsiveSphere} explores structures of repulsive Fermi gases on a spherical surface.

Here we search for the LOFF state in population-imbalanced two-component atomic Fermi gases with pairing interactions on a spherical surface. To incorporate both inhomogeneous effects and spin polarization, we will implement the Bogoliubov-de Gennes (BdG) formalism~\cite{deGennes_book,Zhu-book} and analyze both uniform and inhomogeneous solutions at zero temperature as the spin polarization and pairing interaction change. We will show that while the uniform solutions cease to exist at high spin polarization or low pairing interaction, the LOFF solutions with modulating order parameter and densities emerge in a small regime in the parameter space.

The BdG equation of spin-polarized Fermi superfluid on a spherical surface will show that the number of nodes of the LOFF order parameter increases with population imbalance. In contrast to the two-dimensional (2D) planar case where linear momentum defines the length and energy scales, orbital angular momentum characterizes the properties of the LOFF state on a spherical surface, despite similar equations of state in the uniform, equal-population case~\cite{He22Sphere}. We will see that the LOFF states with odd (even) numbers of nodes in the order parameter are dominated by the lowest few odd (even) spherical harmonics. Meanwhile, local regions with population imbalance emerge whenever the order parameter vanishes at a node, which explains how the LOFF state accommodates spin polarization in real space by spontaneously breaking the spherical symmetry. Furthermore, the compact spherical geometry allows the LOFF states with different numbers of order-parameter nodes to fit in without the complication of incommensurate length scales found in finite rigid lattice systems~\cite{Kim19}, thereby illustrating more geometric effects on many-body physics.

The rest of the paper is organized as follows. Sec.~\ref{Sec:Theory} outlines the BdG formalism for spin-polarized Fermi gases with pairing interactions on a spherical surface. Sec.~\ref{Sec:Solutions} discusses the uniform and LOFF solutions from the BdG equation and their grand potentials for determining the energetically favorable state. Sec.~\ref{Sec:Results} shows a phase diagram of uniform and non-uniform solutions as well as exemplary LOFF solutions near the boundary of the phase diagram. The signature modulating order parameter and density profiles of the LOFF state are also analyzed. Sec.~\ref{Sec:Implications} discusses some implications for probing the LOFF state and future challenges. Finally, Sec.~\ref{Sec:Conclusion} concludes our work.

\section{Theoretical framework}\label{Sec:Theory}
\subsection{Mean-field theory of Fermi gases on a spherical surface}
Here we consider population-imbalanced two-component atomic Fermi gases with contact pairing interactions on a spherical surface. To explore possible phases with different populations of the two components, we implement the BdG equation which is capable of characterizing inhomogeneous Fermi superfluids. We start with the BCS mean field Hamiltonian properly scaled to a unit sphere. This implies the usage of the energy unit $E_0=\hbar^2/(2m_0R^2)$, where $m_0$ and $R$ denoting the mass of the fermions and the radius of the sphere. The Hamiltonian is given by
\be
H_{\mathrm{BCS}}&=&\int_{S^2}d\br\Big[\sum_{\sigma}\psi_{\sigma}^{\dag}(\br)\hat{T}_{\sigma}\psi_{\sigma}(\br)
+\Delta(\br)\psi^{\dag}_{\uparrow}(\br)\psi^{\dag}_{\downarrow}(\br)+ \nonumber \\
& &\Delta^*(\br)\psi_{\downarrow}(\br)\psi_{\uparrow}(\br)\Big].
\label{eq-BCS}
\ee
Here $\br=(\theta,\phi)$ is the spherical coordinate, and we take $\hbar=1=k_B$ throughout the paper. The field operator $\psi_{\sigma}$ is the fermion annihilation operator of spin $\sigma=\uparrow,\downarrow$. The kinematic energy operator is given by
\be
\hat{T}_{\sigma}=-\frac{\nabla^2_s}{2m_0}-\mu_{\sigma}.
\ee
where  $\mu_\uparrow$ and $\mu_\downarrow$ are the chemical potentials of the two components. The spherical Laplacian operator is given by
\be
\nabla_s^2=-\Big(\frac{1}{\sin\theta}\frac{\p}{\p\theta}\sin\theta\frac{\p}{\p\theta}+\frac{1}{\sin^2\theta}\frac{\p^2}{\p^2\phi}\Big).
\ee
In the presence of population imbalance, $\mu_\uparrow\neq\mu_\downarrow$. The gap function, which is also the order parameter, is defined as
\be\label{Eq:gap_def}
\Delta(\br)=-g\ep{\psi_{\downarrow}(\br)\psi_{\uparrow}(\br)},
\ee
where $g$ is the bare coupling constant.

\subsection{BdG formalism}
After introducing the Nambu spinor $\Psi=\Big(\psi_\uparrow(\br),\psi_\downarrow(\br),\psi^\dag_\uparrow(\br),\psi^\dag_\downarrow(\br)\Big)^T$ and discarding a constant term, the BCS mean field Hamiltonian can be rewritten as \cite{deGennes_book,Zhu-book}
\be
H_{\mathrm{BCS}}=\frac12\int_{S^2}d\br\Psi^\dag
\left(
  \begin{array}{cccc}
    T_\uparrow & 0 & 0 & \Delta \\
    0 & T_\downarrow & -\Delta & 0 \\
    0 & -\Delta^* & -T_\uparrow & 0 \\
    \Delta^* & 0 & 0 & -T_\downarrow
  \end{array}
\right)
\Psi^\dag.
\ee
In the absence of spin-orbit coupling or other spin-flipping terms, the above $4\times4$ matrix at each location can be decomposed into two $2\times2$ matrices.
To diagonalize the matrix made by the first and fourth rows and columns, we make the following Bogoliubov transformation 
\be
\psi_\uparrow(\br)=\sum_n u_{1,n}(\br)\beta_{1,n},\quad
\psi^\dag_\downarrow(\br)=\sum_n v_{1,n}(\br)\beta_{1,n}.
\ee
The coefficients $u_{1,n}$ and $v_{1,n}$ then satisfy the BdG equations
\be
\left(
  \begin{array}{cc}
    T_\uparrow & \Delta(\br) \\
    \Delta^*(\br) & -T_\downarrow
  \end{array}
\right)\left(
  \begin{array}{c}
    u_{1,n}(\br) \\
    v_{1,n}(\br)
  \end{array}
\right)=E_n\left(
  \begin{array}{c}
    u_{1,n}(\br) \\
    v_{1,n}(\br)
  \end{array}
\right). \label{eq-BdG1}
\ee
Similarly, for the matrix made by the second and third rows and columns, we introduce the transformation
\be
&&\psi_\downarrow(\br)=\sum_n u_{2,n}(\br)\beta_{2,n},\quad
\psi^\dag_\uparrow(\br)=\sum_n v_{2,n}(\br)\beta_{2,n}.
\ee
Then the coefficients $u_{2,n}$ and $v_{2,n}$ must satisfy the BdG equations 
\be
\left(
  \begin{array}{cc}
    T_\downarrow & -\Delta(\br) \\
    -\Delta^*(\br) & -T_\uparrow
  \end{array}
\right)\left(
  \begin{array}{c}
    u_{2,n}(\br) \\
    v_{2,n}(\br)
  \end{array}
\right)=E_n\left(
  \begin{array}{c}
    u_{2,n}(\br) \\
    v_{2,n}(\br)
  \end{array}
\right)\label{eq-BdG2}
\ee
Due to the particle-hole symmetry of the BdG equation~\cite{Zhu-book}, if $(u_{1,n},v_{1,n})^T$ is an eigenvector of Eq.~(\ref{eq-BdG1}) with eigenvalue $E_n$, then $(v^*_{1,n},-u^*_{1,n})^T$ is an eigenvector of Eq.~(\ref{eq-BdG2}) with eigenvalue $-E_n$. Therefore, we only need to solve Eq.~(\ref{eq-BdG1}) even in the presence of population imbalance. Furthermore, the eigenfunctions $u_{a,n}(\br)$ and $v_{a,n}(\br)$ satisfy the orthogonal condition
\be
\int_{S^2}d\br \Big[u^*_{a,m}(\br)u_{b,n}(\br)+v^*_{a,m}(\br)v_{b,n}(\br)\Big]=\delta_{ab}\delta_{mn}.
\ee

The gap function defined in Eq.~\eqref{Eq:gap_def} is determined by the eigenfunctions through
\be\label{Eq:BdG_gap}
\Delta(\br)=\frac{g}2\sum_{n} u_{1,n}(\br)v^*_{1,n}(\br)\Big[1-2f(E_n)\Big].
\ee
Here $f(x)=\Big[\exp(x/T)+1\Big]^{-1}$ is the Fermi distribution.
%The summation $\sum_n$ is for eigen-energies that satisfy the condition $0\le E_n\le E_{\textrm{cut}}$ with some UV cut off energy $E_{\textrm{cut}}$.
The densities of the two components are given by
\be
&&n_\uparrow(\br)=\sum_n |u_{1,n}(\br)|^2 f(E_n)\label{eq-num1}, \\
&&n_\downarrow(\br)=\sum_n |v_{1,n}(\br)|^2\big[1-f(E_n)\big]. \label{eq-num2}
\ee
We also introduce a fictitious noninteracting two-component equal-population Fermi gas with the same total particle number in the same geometry and use its Fermi energy and momentum $E_F$ and $k_F$ as our units. If $N_\uparrow$ and $N_\downarrow$ denote the fermion numbers of the two components, $ k_F^2/(2m_0)=E_F$ and $E_F=(N_\uparrow+N_\downarrow)/2$ for a continuous dispersion. For the discrete levels from the quantization on a spherical surface, we have $E_F=L_m(L_m+1)$ and $N_\uparrow+N_\downarrow=2(L_m+1)^2$. Here $L_m$ is the highest occupied angular momentum.

\section{Possible solutions}\label{Sec:Solutions}

\subsection{Uniform solutions}
We first explore possible uniform solutions with constant $\Delta$ on the whole sphere. If we expand $u_n(\br)$ and $v_n(\br)$ by the spherical harmonics $Y_{lm}(\theta,\phi)$, the BdG equation becomes an algebraic equation for given quantum numbers $l$ and $m$ as
\be
\left(
  \begin{array}{cc}
    (\xi_l-h) & \Delta \\
    \Delta & -(\xi_l+h)
  \end{array}
\right)\left(
  \begin{array}{c}
    U \\
    V
  \end{array}
\right)=E\left(
  \begin{array}{c}
    U \\
    V
  \end{array}
\right),
\ee
where $\xi_l=l(l+1)/(2m_0)-\mu$, $\mu=(\mu_\uparrow+\mu_\downarrow)/2$ and $h=(\mu_\uparrow-\mu_\downarrow)/2$.
The eigenvalues of the above equation are $E=\pm E_l-h$ with the corresponding eigenvectors
\be
&&(U,\,V)=(u_l,\,v_l),\quad (-v_l,\,u_l);\\
&&u_l^2=\frac12\Big(1+\frac{\xi_l}{E_l}\Big),\quad
v_l^2=\frac12\Big(1-\frac{\xi_l}{E_l}\Big),
\ee
where $E_l=\sqrt{\xi_l^2+\Delta^2}$.

With the above results, we find the number equations 
\be
&&n_\uparrow=\sum_l(2l+1)\Big[u_l^2f(E_{l,\uparrow})+v_l^2\big(1-f(E_{l,\downarrow})\big)\Big],\\
&&n_\downarrow=\sum_l(2l+1)\Big[u_l^2f(E_{l,\downarrow})+v_l^2\big(1-f(E_{l,\uparrow})\big)\Big].
\ee
We also introduce $n=n_\uparrow+n_\downarrow$ and $\delta n=n_\uparrow-n_\downarrow$. 
%we obtain
%\be
%&&n=\sum_l(2l+1)\Big[1-\frac{\xi_l}{E_l}+[f(E_{l,\uparrow})+f(E_{l,\downarrow})]\frac{\xi_l}{E_l}\Big], \\
%&&\delta n=\sum_l(2l+1)\Big[f(E_{l,\downarrow})-f(E_{l,\uparrow})\Big].
%\ee
Here $E_{l,\uparrow}=E_l-h$ and $E_{l,\downarrow}=E_l+h$. Moreover, the spin polarization is defined as
\be\label{Eq:spol}
p=\frac{n_\uparrow-n_\downarrow}{n_\uparrow+n_\downarrow}.
\ee

Similarly, we find that the gap equation 
\be
\frac1g=\sum_l(2l+1)\frac{1-f(E_{l,\uparrow})-f(E_{l,\downarrow})}{2E_l}.
\ee
The bare coupling $g$ can be regularized as
\be
\frac 1g=\sum_l \frac{2l+1}{2\epsilon_l+|\epsilon_b|}.\label{eq-eb}
\ee
Here $\epsilon_l=l(l+1)/(2m_0)$, and $\epsilon_b$ is the two-body binding energy related to the two-body scattering length $a$ via
$\epsilon_b=-\dfrac{1}{ma^2}$. The interaction strength will be indicated by the dimensionless quantity $-\ln(k_Fa)$ in the following.
%The Fermi energy of an equal-population Fermi gas with the same total fermion number is given by $E_F=L_m(L_m+1)$ where $L_m$ is the highest occupied angular momentum determined by $N=2(L_m+1)^2$. Then the Fermi momentum is given by $k_F^2/(2m_0)=E_F$.
To find out the scattering length, we can compute the binding energy $\epsilon_b$ from the bare coupling $g$ through Eq.~(\ref{eq-eb}). Then, the scattering length $a$ can be obtained from $\epsilon_b$. %Again, the interaction strength is indicated by the dimensionless quantity $-\ln(k_Fa)$.

%To find out the two-body scattering length, we make use the following relation
%\be
%\frac 1g=\int dl\frac{2l+1}{2\epsilon_l+|\epsilon_b|}=\int_0^{E_{\textrm{cut}}}\frac{d\epsilon_l}{2\epsilon_l+|\epsilon_b|}.\\
%&&E_b=-\frac{1}{m_0 a^2}.
%\ee
%\textbf{Please explain $E_{cut}$.}

\subsection{LOFF solutions}
We seek inhomogeneous solutions similar to the LOFF superfluid of the BdG equation due to population imbalance. On a spherical surface, the quasi-particle wavefunctions $u_n(\br)$ and $v_n(\br)$ are expanded by the spherical harmonics as
\be
&&u_{1,n}(\br)=\sum_{l,m}c_{nlm}Y_{l,m}(\theta,\phi), \\
&&v_{1,n}(\br)=\sum_{l,m}d_{nlm}Y_{l,m}(\theta,\phi).
\ee
The expansion transforms the BdG equations into a large matrix equation. Hereafter we will look for solutions with azimuthal symmetry, so the magnetic quantum number $m$ is a good quantum number. Therefore, the spherical harmonics with different $m$ are decoupled. Then the matrix form of BdG equation splits into smaller diagonal blocks where each block corresponds to different values of $m$. 
For a given $m$, the BdG equation can be written as
\be\label{Eq:BdG}
&&\sum_{ll'}\left(
  \begin{array}{cc}
    T_{\uparrow,m} & D_m \\
    D_m^T & -T_{\downarrow,m}
  \end{array}
\right)_{ll'}\left(
  \begin{array}{c}
    c_{nl'm} \\
    d_{nl'm}
  \end{array}
\right)=E^{(m)}_n\left(
  \begin{array}{c}
    c_{nlm} \\
    d_{nlm}
  \end{array}
\right), \nonumber\\
\ee
where $l=m,m+1,\cdots$ and $s=\uparrow,\downarrow$, $(T_{s,m})_{ll'}=\Big(\frac{l(l+1)}{2m_0}-\mu_s\Big)\delta_{ll'}$, and 
$(D_m)_{ll'}=\int_{-1}^1dx\Delta(x)N_{lm}P^m_l(x)N_{l',m+1}P^{m}_{l'}(x)$.
Here $m_0$ is the mass of the fermion, not to be confused with the quantum number $m$, $P_l^m(x)$ is the associated Legendre polynomial, and $x=\cos\theta$. The normalization factor is $N_{lm}=\sqrt{\frac{(l-|m|)!}{(l+|m|)!}\frac{2l+1}{2}}$.
In principle, a differential operator will be transformed into a matrix of infinite dimensions. In numerical practice, we have to truncate the operator into a matrix with finite dimensions. In our numerical calculation, we expand the wave-function up to the largest angular quantum number $l_{max}=30$. We have checked that the variation of the cutoff of $l_{max}$ will not affect the qualitative features of the Fermi superfluid.

Numerically, we use the iterative method~\cite{Zhu-book} to solve the BdG equation by starting with a trial function for $\Delta$, solving the eigenvalue problem~\eqref{Eq:BdG}, assembling a new $\Delta$ according to Eq.~\eqref{Eq:BdG_gap}, and repeating the procedure until the convergence condition $\int d\mathbf{r} |\Delta_{new}-\Delta_{old}|\le \epsilon$ is satisfied. In our numerical calculations, we use $\epsilon=10^{-3}$. 
In our search for the LOFF solutions, we make an initial guess of an inhomogeneous gap function. For example, a trial initial gap function of form 
$\Delta(x)=\Delta_0Y_{10}(\theta,\phi)
=\Delta_0\sqrt{\frac{3}{4\pi}}\cos\theta$, which linearly depends on $x$, may be used to start the iterative solver for the BdG equation.

\subsection{Grand potential}
To check if the LOFF state is energetically more favorable than the uniform solution, we also compare the grand potential, given by $E=\langle H-\mu N\rangle-TS$~\cite{Fetter_book}, in the presence of population imbalanced when both LOFF and uniform solutions are available.
At $T=0$, the grand potential of the uniform solution of polarized superfluid can be obtained by generalizing the formalism of Ref.~\cite{Tinkham}. Explicitly, 
\be
E_{uni}&=&\frac{\Delta^2}{g}+\sum_l(2l+1)(\xi_l-E_l)\label{eq-Euni}\\
&=&\sum_l(2l+1)\Big(\xi_l-E_l-\frac{\Delta^2}{2E_l}\Big).\nonumber
\ee
In the second line, we have made use of the gap equation to handle the bare coupling $g$.

Similarly, the grand potential of the LOFF state is given by
\be
E_{LOFF}&=&\frac{1}{4\pi g}\int_{S^2}d\br\Delta(\br)+\frac12\sum_m\Big[\mathrm{Tr}(T_{\uparrow,m}+T_{\downarrow,m})- \nonumber \\
&&\sum_n|E^{(m)}_n|\Big].
\ee
One can check that the above formula reduces to Eq.~(\ref{eq-Euni}) when the gap function is uniform. We caution that while the configuration with a lower grand potential is more stable, the system may stay in the metastable configuration if the preparation or condition does not allow proper relaxation into the stable state.

\section{Numerical results}\label{Sec:Results}
For given values of $\mu_{\uparrow,\downarrow}$, Eqs.~(\ref{eq-num1}) and (\ref{eq-num2}) are used to determine the fermion numbers $N_\uparrow$ and $N_\downarrow$. After determining the Fermi energy $E_F$ and wave vector $k_F$ of the corresponding noninteracting systems with the same total particle number, the dimensionless quantities with respect to $E_F$ and $k_F$ are presented. The spin polarization is computed according to Eq.~\eqref{Eq:spol}.
%\be
%p=\frac{N_1-N_2}{N_1+N_2}.
%\ee

\begin{figure}[t]
\centering
\includegraphics[width=0.8\columnwidth]{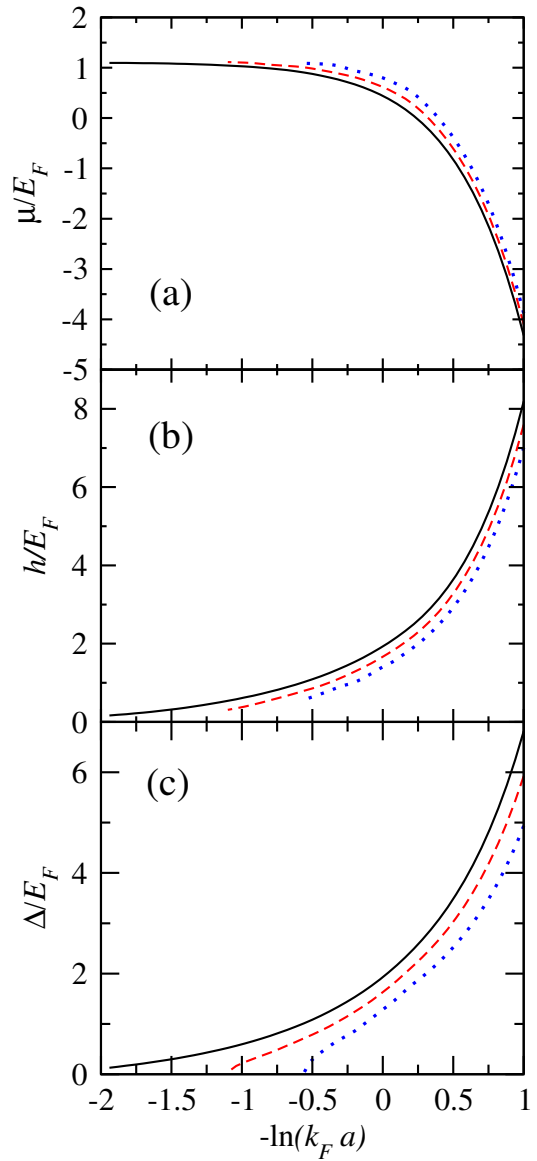}
\caption{Uniform solutions showing (a) $\mu$, (b) $h$, and (c) $\Delta$ as functions of $-\ln(k_Fa)$ at $T=0$. The solid, dash, and dotted lines are for $p=0.1$, $0.3$, and $0.5$, respectively. No uniform solution can be found beyond the point when $\Delta\rightarrow 0$.
}
\label{mu-T0}
\end{figure}

\subsection{Uniform solutions and phase diagram}
Fig. \ref{mu-T0} shows $\mu$, $h$ and $\Delta$ of the uniform solutions as functions of $-\ln(k_Fa)$ at $T=0$ for $p=0.1, 0.3, 0.5$, respectively. The monotonic decreasing of $\mu$ and increasing of $\Delta$ as the pairing strength increases are similar to the equal-population case. The BCS-BEC crossover may be identified by the point where $\mu$ changes sign, which is around $-\ln(k_Fa)\approx 0$ for our setup. 
As shown in Fig.~\ref{mu-T0}(c), $\Delta\rightarrow 0$ for population-imbalanced Fermi gases as $-\ln(k_Fa)$ decreases into the BCS regime. The vanishing order parameter is consistent with the Clogston-Chandrasekhar limit~\cite{PhysRevLett.9.266,10.1063/1.1777362} as the uniform BCS Fermi superfluid is no longer a feasible solution in the presence of population imbalance. Beyond the $\Delta\rightarrow 0$ point, inhomogeneous solutions are expected to emerge.

\begin{figure}[t]
\centering
\includegraphics[width=0.8\columnwidth]{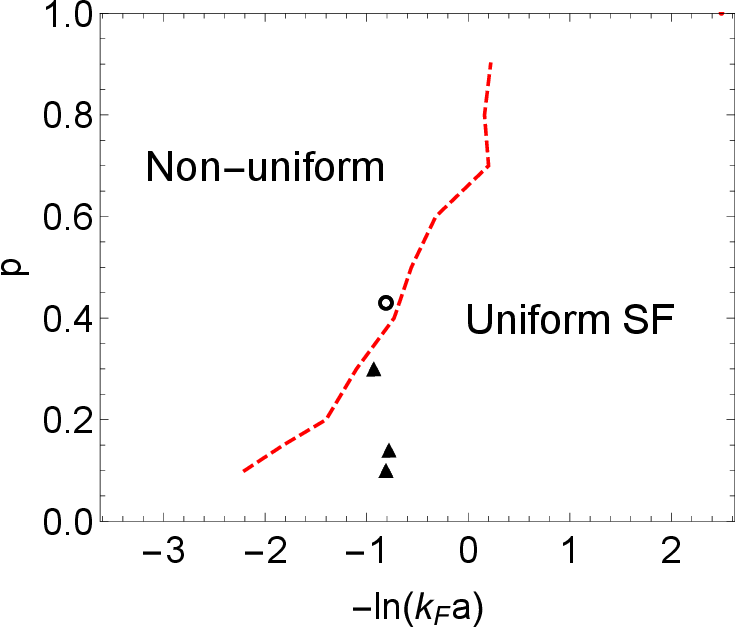}
\caption{Phase diagram of spin polarized Fermi superfluid on a spherical surface at $T=0$. Here ``Uniform SF'' labels the region where uniform spin-polarized superfluid solutions exist. ``Non-uniform'' labels the region beyond uniform solutions, presumably occupied by phase separation or LOFF state. The boundary is determined by estimating where the uniform solutions first give a vanishing order parameter. The three triangles correspond to the LOFF solutions in Fig. \ref{LO-all} while the open circle corresponds to the LOFF solution in Fig. \ref{LO-nonuniform}. 
}
\label{phase}
\end{figure}

Fig.~\ref{phase} shows a phase diagram according to where the uniform solution ceases to exist due to the vanishing order parameter. The uniform solutions occupy the regime with stronger pairing interactions and lower population imbalance. To facilitate a fair comparison with the BdG equation for inhomogeneous calculations presented later, we set $l_{max}=30$ when performing the summations in the equations of state of the uniform cases rather than approximating the summation by an integral.
The kinky boundary in Fig.~\ref{phase} is due to the finite summation over $l$.
%When $\Delta\rightarrow 0$ in our uniform solutions, we relax the homogeneity condition and look into inhomogeneous solutions from the BdG equation.  

\begin{figure}[t]
\centering
\includegraphics[width=\columnwidth]{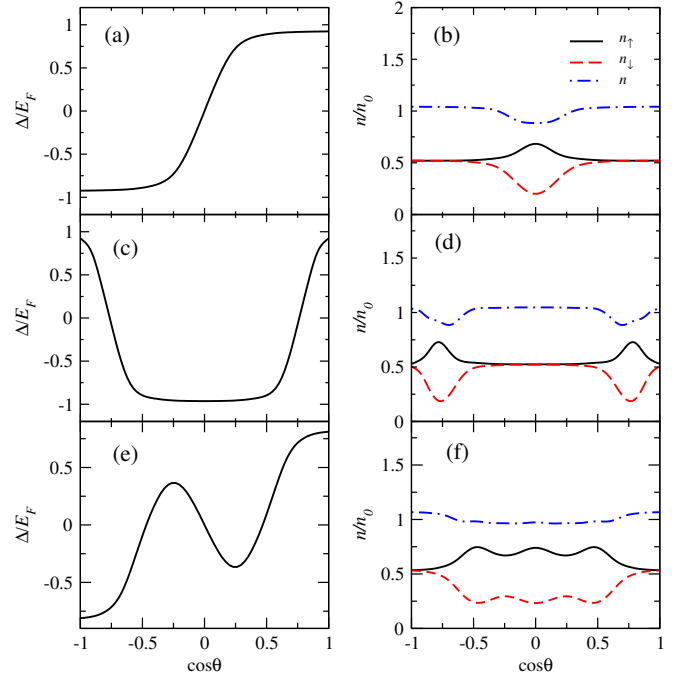}
\caption{Panels (a),(c),(e): Order parameter $\Delta/E_F$ as a function of $\cos\theta$. Panels (b),(d),(f): Spin-resolved densities $n_\uparrow/n_0$ (black solid lines), $n_\downarrow/n_0$ (red dashed lines), and total density $n/n_0$ (blue dot-dash lines) as functions of $\cos\theta$, where $n_0=N/(4\pi)$. For (a) and (b), $-\ln(k_Fa)=-0.81$, $p=0.1$ and $\mu_\uparrow/E_F=1.4$, $\mu_\downarrow/E_F=0.56$. For (c) and (d), $-\ln(k_Fa)=-0.78$ and $p=0.14$, $\mu_{\uparrow,\downarrow}$ are the same as (a) and (b). For (e) and (f), $-\ln(k_Fa)=-0.93$, $p=0.3$ and $\mu_\uparrow/E_F=1.6$, $\mu_\downarrow/E_F=0.53$. For all cases, $T=0$.
}
\label{LO-all}
\end{figure}

\subsection{LOFF solutions}
\subsubsection{Within the uniform-solution regime}
Overlapping with the regime where uniform solutions can be found but near the phase boundary, the LOFF solutions can be obtained from the BdG equation. Fig. \ref{LO-all} shows some selected results of the LOFF solutions with different pairing interactions and population imbalance. The left column shows how $\Delta$ behaves on the spherical surface, featuring the spatial modulations due to population imbalance. As the spin polarization increases, the order parameter develops more nodes and indicates more rapid oscillations on the spherical surface.

\begin{figure}[t]
\centering
\includegraphics[width=0.7\columnwidth]{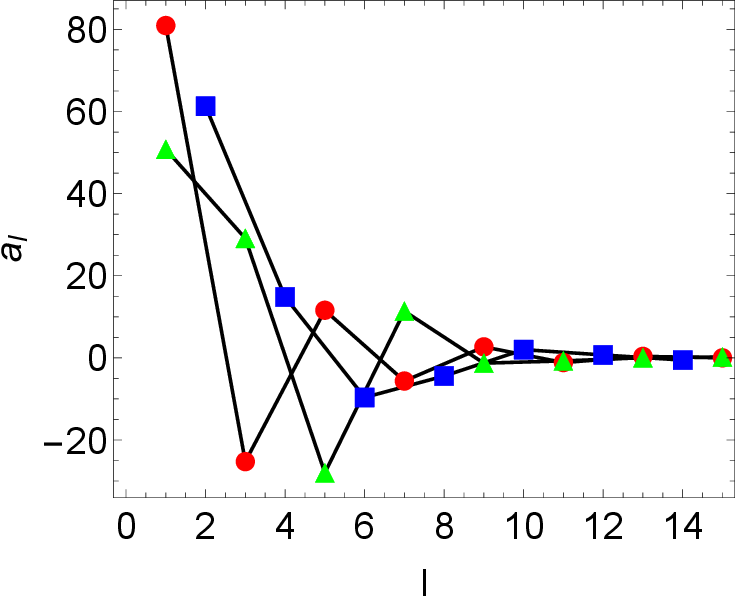}
\caption{The expansion coefficients $a_l$ of the LOFF order parameter according to Eq.~(\ref{eq-del-ex}) as a function of $l$. The red circles, blue squares, and green triangles correspond to the one-, two-, and three-node LOFF states of Fig. \ref{LO-all}, respectively.
}
\label{del-ex}
\end{figure}

However, the convergent solutions of $\Delta$ from the BdG equation are no longer simple functions of $\cos\theta$, despite their relatively plain appearance. To determine the composition of the order parameter of the LOFF states, we expand the convergent solution $\Delta$ of the BdG equation by the Legendre polynomial, 
\be
\Delta(x)=\sum_{l=0}^{\infty}a_l\sqrt{\frac{2l+1}{2}}P_l(x).\label{eq-del-ex}
\ee
When $\Delta(x)$ is an odd (even) function of $\cos\theta$, the coefficient $a_l$ is non-zero only if $l$ is an odd (even) integer. We plot the nonzero $a_l$ up to $l=25$ in Fig. \ref{del-ex} for the three lowest modulating LOFF solutions of Fig.~\ref{LO-all} with $1,2,3$ nodes in the order parameter. One can see that the most important contributions come form the low angular momentum components while higher-order coefficients are almost zero. Therefore, the fully convergent $\Delta$ reveals contributions from various spherical harmonics to the LOFF order parameter on a spherical surface.

The spin-resolved densities are shown in the right column of Fig. \ref{LO-all}. For the LOFF state with a single node in $\Delta$, there is a small bump in $n_\uparrow$ and a dip in $n_\downarrow$ around the equator ($\cos\theta=0$), showing where population imbalance emerges on the sphere. In contrast to the uniform solutions with homogeneous density profiles, the total fermion density $n(\br)=n_\uparrow(\br)+n_\downarrow(\br)$ is not uniform. Interestingly, the total density is reduced when the population imbalance is maximal.

By increasing the spin polarization, more population-imbalanced regions of the LOFF solutions appear on the spherical surface. Importantly, the locations of the population-imbalanced regions coincide with the nodes of the order parameter. Since the suppression of the order parameter around the nodes reduces pairing, the system can accommodate spin polarization locally. Meanwhile, the presence of those population-imbalanced regions result in local minima of the total density. Therefore, the profiles of spin-resolved and total densities reveal the oscillatory behavior of the LOFF state and help identify the nodes of the order parameter.

\begin{figure}[t]
\centering
\includegraphics[width=0.7\columnwidth]{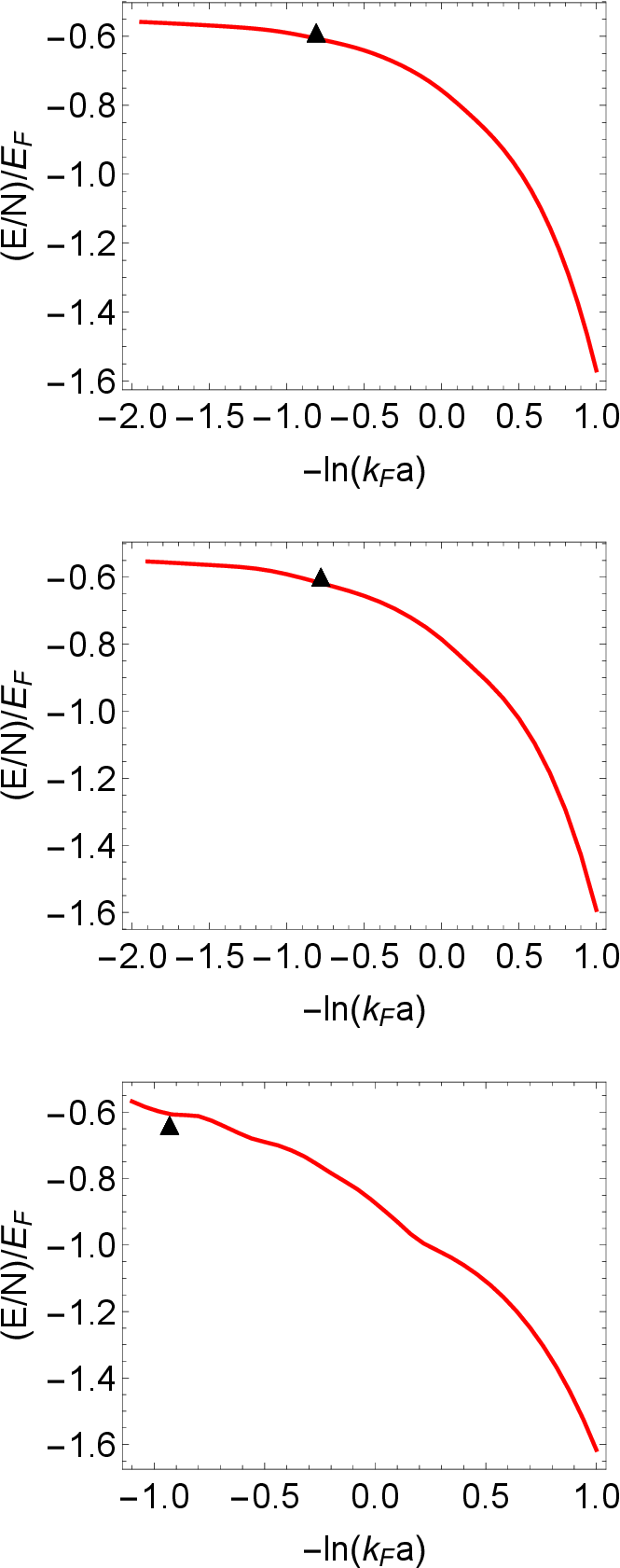}
\caption{The normalized grand potential per particle $(E/N)/E_F$ of spin-polarized superfluid as a function of $-\ln(k_Fa)$ at $T=0$. Here $p=0.1, 0.14, 0.3$ from top to bottom, which correspond to the cases shown in Fig.~\ref{LO-all}. The red lines represent the uniform states. The black triangles are selected LOFF states with one, two, and three nodes in the order parameter when the polarization increases. Only the three-node LOFF state in the bottom is energetically more favorable than the uniform solution.
}
\label{E-ka}
\end{figure}

\begin{figure}[t]
\centering
\includegraphics[width=\columnwidth]{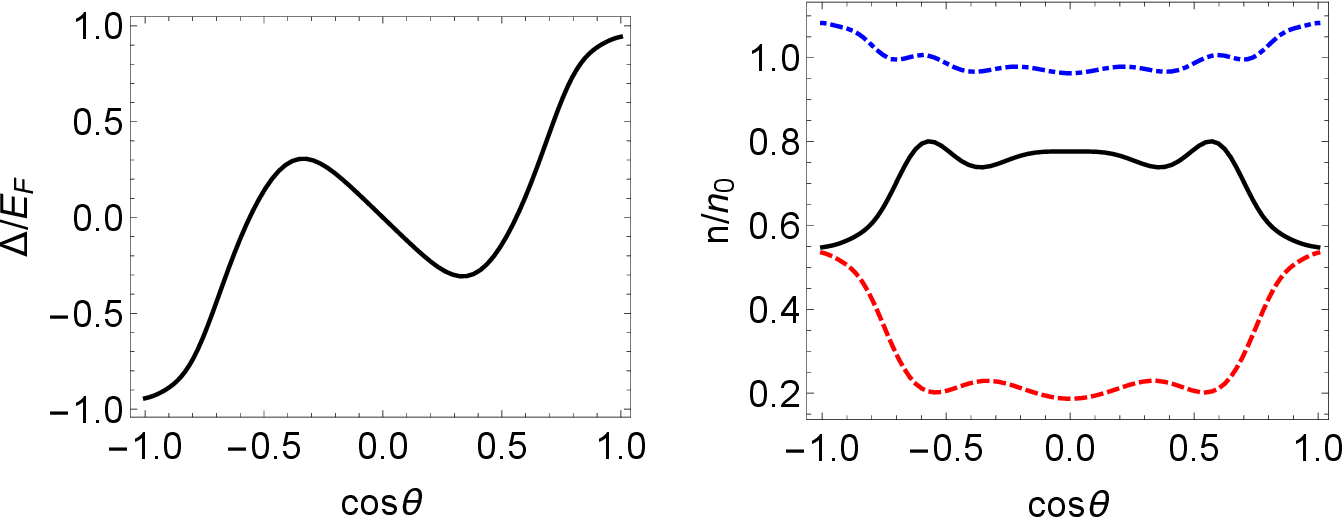}
\caption{(Left) The order parameter $\Delta/E_F$ as a function of $\cos\theta$. (Right) The spin-resolved densities $n_\uparrow/n_0$ (black solid line),  $n_\downarrow/n_0$ (red dashed line), and the total density $n/n_0$ (blue dot-dash line) as functions of $\cos\theta$, where $n_0=N/(4\pi)$. Here $-\ln(k_Fa)=-0.81$, $p=0.43$, $\mu_\uparrow/E_F=1.78$, $\mu_\downarrow/E_F=0.35$, and $T=0$.
}
\label{LO-nonuniform}
\end{figure}

To determine whether the LOFF solution is more energetically favorable than the corresponding uniform solution, we evaluate the grand potentials of the solutions with comparable spin polarization and selected pairing interactions from the BdG equation.
In Fig. \ref{E-ka}, we plot the grand potential per particle $E/N$ of the uniform solutions of population-imbalanced superfluids as a function of the pairing interaction. The grand potential decreases with the pairing interaction due to the stronger condensation energy~\cite{Fetter_book}. We note that for a 2D free Fermi gas on a plane, the ground state energy per particle is $0.5E_F$. If we subtract the chemical potential $\mu=E_F$, the grand potential per particle of the free Fermi gas measured from the Fermi surface is $-0.5E_F$. From Fig. \ref{E-ka}, one can see that the grand potential per particle of the uniform solution in the BCS limit is slightly lower than that of the free Fermi gas, which correctly reflects the condensation energy of the superfluid~\cite{Fetter_book}. 

Meanwhile, the black triangles in Fig. \ref{E-ka} indicate the grand potentials of the LOFF states shown in Fig.~\ref{LO-all} with one, two, and three nodes in the order parameter from the BdG equation as the polarization increases. While the LOFF states with lower polarization and fewer nodes have slightly higher grand potentials relative to those of the uniform solutions, the three-node LOFF state with higher polarization has a lower grand potential than the uniform solution. 
Therefore, near the boundary but within the regime of the uniform solutions shown in Fig.~\ref{phase}, the LOFF states with lower node counts may be preempted by the corresponding uniform solutions when the polarization is low. This is understandable because the LOFF modulations increases the kinetic energy. Nevertheless, the LOFF states with higher node counts may compete with the uniform solution at higher spin polarization and become energetically favorable since the uniform solutions need to adjust the dispersions more drastically to accommodate the polarization uniformly while the LOFF state allows modulating population imbalance in real space. 
%Beyond the boundary of Fig.~\ref{phase} when the uniform solutions are no longer possible from the BCS formalism, the BdG equation typically produces the LOFF solutions for polarized Fermi gases on a spherical surface.

We remark that a precise comparison between a LOFF state and a uniform state is complicated by the different spin-resolved densities in the convergent BdG solutions, making an exact match of the polarization and chemical potentials extremely challenging numerically. Nevertheless, the general trend is that when a uniform solution is available at low spin polarization, the corresponding LOFF solution with few nodes in the order parameter may not be more energetically favorable. As the polarization increases, the uniform solution may no longer provide suitable accommodation for homogeneous population imbalance through its adjustment of energy dispersions, but the LOFF state with a higher node count in its order parameter can accommodate density differences around its nodes. Therefore, the LOFF solution emerges as the more stable solution near the boundary of Fig.~\ref{phase} at higher polarization  when the uniform solutions are still available.

\subsubsection{Beyond the uniform-solution regime}
In the regime where uniform solutions are no longer possible in Fig.~\ref{phase}, the default solutions are phase separation between BCS superfluid and polarized normal phase similar to 3D population imbalanced Fermi gases~\cite{Chien2007a,Parish2007}. We caution that phase separation with a BCS phase occupying one part of the sphere and a polarized normal phase occupying another does not come out of the BdG equation as a convergent solution, no matter what initial conditions we provide. One may resort to manually matching the two phases and their grand potentials as performed in previous studies of 3D cases~\cite{Chien2007a}. Nevertheless, LOFF phases with higher counts of nodes in the order parameter can still be found from the BdG equation near the boundary. Fig.~\ref{LO-nonuniform} shows the profiles of the order parameter and densities of a three-node LOFF phase found at the open-circle symbol shown in the non-uniform regime of Fig.~\ref{phase}.

By comparing Fig.~\ref{LO-all} (e) and (f) with Fig.~\ref{LO-nonuniform}, one can see that the LOFF states have similar profiles of the order parameter and densities in both regimes of Fig.~\ref{phase}. However, the phase boundary of Fig.~\ref{phase} extends drastically towards the low-interaction regime as the spin-polarization decreases. Consequently, our numerical calculations of the BdG equation do not generate fully convergent LOFF solutions at low spin polarization beyond the uniform-solution regime, indicating the limited availability of the LOFF solution in the parameter space. Therefore, the LOFF phase is relatively rare on a 2D spherical surface according to the BdG equation, similar to the 3D case where the LOFF phase occupies a very limited regime in the parameter space~\cite{Parish2007} as well. This is in contrast to the (quasi) 1D case, where the LOFF state is predicted to be more robust~\cite{PhysRevResearch.3.043105}.

\section{Implications}\label{Sec:Implications}
While uniform and LOFF polarized Fermi superfluids on spherical surfaces are obtained from the mean-field theory and BdG equations, early experiments on population-imbalanced atomic Fermi gases with pairing interactions in 3D harmonic traps~\cite{Zwierlein2006,Partridge2006} indicate large regimes in the parameter space are phase separation between an nearly unpolarized BCS Fermi superfluid and a spin-polarized normal gas. In our BdG calculations, however, we have not found credible phase-separation structures between the BCS and normal phases, mainly due to the repeated oscillatory behavior in the intermediate steps. While low-dimensional Fermi gases with pairing interactions seem to favor the LOFF state in  a larger regime~\cite{Liao2010,PhysRevResearch.3.043105}, the thin-shell structure of spin-polarized atomic superfluid studied here offers a promising platform for realizing and characterizing the elusive LOFF state in compact and curved geometry. However, the LOFF states on a spherical surface were only found in a narrow regime close to the end of the uniform solution, and only higher-node LOFF states are favorable at higher spin polarization, making the realization still a challenge in future experiments.

The LOFF state features modulating order parameter and spin-resolved densities, as shown in Figs.~\ref{LO-all} and \ref{LO-nonuniform}. The densities of different components of cold-atom systems can be measured by spin-resolved imaging~\cite{PhysRevA.97.023410,PhysRevA.77.033401}. Therefore, local population-imbalanced regions can be identified from the spin-resolved density profiles, which then indicate the nodes of the order parameter in the LOFF phase. A direct measurement of the oscillating order parameter of the LOFF phase remains more challenging. However, spatially resolved radio frequency spectroscopy~\cite{PhysRevLett.99.090403,doi:10.1126/science.aan5950} poses to be a powerful tool for revealing the modulating gap function due to the LOFF order parameter.

So far, we focus on the ground-state properties of population-imbalanced Fermi gases on a spherical surface. As explained in Ref.~\cite{He22Sphere}, finite-temperature effects can result in a complicated phase diagram of spherical Fermi gases with pairing interactions even when the populations are equal. This is because a spherical surface is 2D, indicating that topological excitations like quantum vortices can be activated by thermal effects~\cite{Berezinskii71,Berezinskii72,KT73}. Moreover, pairing-fluctuation effects~\cite{HePRA15,Levin2D,Enss16} in 2D Fermi superfluid also become prominent when the pairing strength and temperature increase. The BdG formalism requires substantial modifications if those finite-temperature effects are considered, leaving finite-temperature spin-polarized Fermi gases with pairing interactions on a spherical surface a challenge for future theoretical and experimental research.

\section{Conclusion}\label{Sec:Conclusion}
The above analysis of atomic Fermi gases with pairing interaction and population imbalance on a spherical surface by the BdG formalism shows that possible LOFF states with modulating order parameters and density profiles can emerge as a ground-state candidate when the uniform solutions are suppressed by spin polarization. The number of nodes in the LOFF order parameter increases with population imbalance, and the density difference is prominent around the order-parameter nodes to accommodate the excess fermions in real space. The LOFF states are only found near the boundary where uniform solutions are no longer possible, thereby more akin to the fragile 3D LOFF cases than the robust 1D cases. Although the spherical surface imposes periodic boundary condition in arbitrary directions, the LOFF solutions spontaneously break the spherical symmetry and exhibit its signature oscillatory behavior in real space. The spherical-shell geometry for ultracold atoms thus provides an interesting platform for characterizing inspiring many-body phenomena, including the elusive LOFF state.

\begin{acknowledgments}
Y. H. was supported by the National Natural Science Foundation of China under Grant No. 11874272. C.-C. C. was partly supported by the DOE (No. DE-SC0025809).
\end{acknowledgments}

%\begin{thebibliography}{99}

%\bibitem{Fetter}
%A. L. Fetter and J. D. Walecka, \textit{Quantum Theory of Many-Particle Systems}, McGraw-Hill, San Francisco, 1971.

%\end{thebibliography}

%\bibliography{reference}

\end{document}